\documentclass[fleqn,usenatbib]{mnras}

\usepackage{newtxtext,newtxmath}

\usepackage[T1]{fontenc}
\usepackage{ae,aecompl}
\usepackage{multirow}

\usepackage{graphicx}	
\usepackage{amsmath}	
\usepackage{amssymb}	

\def\aj{AJ}%
\def\araa{ARA\&A}%
\def\apj{ApJ}%
\def\apjl{ApJ}%
\def\apjs{ApJS}%
\def\ao{Appl.~Opt.}%
\def\aap{A\&A}%
\def\mnras{MNRAS}%
\def\nat{Nature}%
\def\jgr{J.~Geophys.~Res.}%
\title[The ALMA view of TWA\,7]{Sub-millimeter non-contaminated detection of the disk around TWA\,7 by ALMA.}

\author[A. Bayo et al.]{
A. Bayo$^{1,2}$,\thanks{E-mail: amelia.bayo@uv.cl}
J. Olofsson$^{1,2}$,
L. Matr\`a$^{3}$
J. C. Beam\'in$^{1,2}$,
J. Gallardo$^{4}$,
\newauthor I. de Gregorio-Monsalvo$^{4,2}$,
M. Booth$^{5}$,
C. Zamora$^{1,2}$, 
D. Iglesias$^{1,2}$,
Th. Henning$^{6}$, 
\newauthor M. R. Schreiber$^{1,2}$ and
C. C\'aceres$^{7,2}$
\\
$^{1}$Instituto de F\'isica y Astronom\'ia, Facultad de Ciencias, Universidad de Valpara\'iso, Av. Gran Breta\~na 1111, Valpara\'iso, Chile\\
$^{2}$N\'ucleo Milenio Formaci\'on Planetaria - NPF, Universidad de Valpara\'iso, Av. Gran Breta\~na 1111, Valpara\'iso, Chile\\
$^{3}$Harvard-Smithsonian Center for Astrophysics, 60 Garden Street, Cambridge, MA 02138, USA\\
$^{4}$Atacama Large Millimeter / Submillimeter Array, Joint ALMA Observatory, Alonso de C\'ordova 3107, Vitacura 763-0355, Santiago,
Chile\\
$^{5}$Astrophysikalisches Institut und Universit\"atssternwarte, Friedrich-Schiller-Universit\"at Jena, Schillerg\"a\ss{}chen 2-3, 07745 Jena, Germany\\
$^{6}$Max Planck Institut f\"ur Astronomie, K\"onigstuhl 17, D-69117 Heidelberg, Germany\\
$^{7}$Departamento de Ciencias Fisicas, Facultad de Ciencias Exactas, Universidad Andres Bello. Av. Fernandez Concha 700, Las Condes, Santiago, Chile.
}

\date{Accepted XXX. Received YYY; in original form ZZZ}

\pubyear{2015}

\begin{document}
\label{firstpage}
\pagerange{\pageref{firstpage}--\pageref{lastpage}}
\maketitle

\begin{abstract}
Debris disks can be seen as the left-overs of giant planet formation and the possible nurseries of rocky planets. While M-type stars out-number more massive stars we know very little about the time evolution of their circumstellar disks at ages older than $\sim 10$\,Myr. Sub-millimeter observations are best to provide first order estimates of the available mass reservoir and thus better constrain the evolution of such disks. Here, we present ALMA Cycle\,3 Band\,7 observations of the debris disk around the M2 star TWA\,7, which had been postulated to harbor two spatially separated dust belts, based on unresolved far-infrared and sub-millimeter data. We show that most of the emission at wavelengths longer than $\sim 300$\,$\mu$m is in fact arising from a contaminant source, most likely a sub-mm galaxy, located at about $6.6\arcsec$ East of TWA\,7 (in 2016). Fortunately, the high resolution of our ALMA data allows us to disentangle the contaminant emission from that of the disc and report a significant detection of the disk in the sub-millimeter for the first time with a flux density of 2.1$\pm$0.4 mJy at 870 $\micron$. With this detection, we show that the SED can be reproduced with a single dust belt.
\end{abstract}



\section{Introduction}

Circumstellar disks are natural consequences of the star formation process; as the initial core collapses, gas and dust rotating around the central mass will fall towards the plane of rotation. At these early stages the gas-to-dust mass ratio is usually assumed to be identical to the ISM ($\sim 100$), thus the mass of the disk is dominated by the gas, while its opacity (and therefore its emission at long wavelengths) is dominated by the dust (\citealp{Henning2010}). With a half-life time of a few million years (\citealp{Haisch2001}), the gas-rich circumstellar disks rapidly evolve dissipating most of their gas via mechanisms such as photo-evaporation (\citealp{Alexander2006}).

When the disk enters its ``debris disk'' phase (at a canonical age of $\sim 10$\,Myr, see the review by \citealp{Matthews2014}) only dust grains, planetesimals and possibly already formed giant planets will remain. 
In the earliest phases ($10-100$ Myr), it is thought that collisions among planetesimals can lead to the formation of a few oligarchs that may later on form rocky terrestrial planets (\citealp{Kenyon2008}, but see also \citealt{Johansen15}, for instance, for faster mechanisms such as streaming instability). Afterward, the evolution of the entire disk becomes much more monotonic as the larger bodies are ground down in a collisional cascade \citep{Dohnanyi1969}. Removal processes such as radiation pressure (for intermediate and solar type stars), or stellar winds (for late-type stars) are very efficient at removing the low end of the grain size distribution. Over time, the mass reservoir is slowly depleted from the disk (following $t^{-1/2}$ according to \citealp{Holland2017}).

Interestingly, M-type stars constitute about $70$\% of the total number of stars, and are gaining popularity in the search of planetary systems given the favorable contrast in terms of size and brightness that they can offer. This is in line with the analysis of \textit{Kepler} observations presented in \citet{Dressing2015}, who reported a planet occurrence rate of about $2.5$ planets (with radius of $1-4$\,R$_\oplus$ and periods shorter than $200$\,days) for M-type stars. This suggests that rocky planet formation (at least) is efficient around low-mass stars. However, our understanding of the cradles of these possible planets, the debris disks around low-mass stars, is strongly limited by the small number of objects that are known to harbor such disks. Whilst there might be some fundamental difference about M-type stars that make their discs very different to those around earlier spectral types (see for example the extreme case of the ``hybrid" disk around the 45Myr old accreting WISE080822, \citealt{Murphy17}), it is also possible that this desert of debris disks is related to an observational sensitivity limit \citep{Morey2014}.

Consequently, little is known about debris disks around low-mass stars, especially when it comes to spatially resolved observations. As a matter of fact, two of the very few debris disks with near-IR high angular resolution observations show intriguing features. 
Fast-moving structures have been detected in the edge-on disk around AU\,Mic \citep{Boccaletti2015,Boccaletti2018}, and a faint spiral arm and outer disk have been reported in the disk around TWA\,7 \citep{Olofsson2018}. The other debris disks with spatially resolved observations being TWA\,25 \citep[][who also presented the first image of TWA\,7]{Choquet2016}, GJ\,581 \citep{Lestrade2012}, and GSC\,07396-00759 \citep{Sissa2018}.
Scattered light observations provide invaluable information about the spatial distribution of the small dust grains, at exquisite angular resolution, but are often not sufficient to reach a comprehensive characterization of the entire system. Spatially resolved (sub-)\,mm observations, with ALMA for instance, are highly complementary as they trace larger grains ($100$\,$\mu$m to $\sim 1$\,mm) located in the ``birth ring''. Furthermore, they provide us with a first order estimate of the total dust mass in the disk. There is therefore a great synergy in multi-wavelength observations and analysis. 

TWA\,7 is an M2-type star belonging to the TW Hydrae association (TWA, $7.5\pm0.7$\,Myr old, \citealt{Ducourant2014}), making it
one of the closest ($34.029 \pm 0.076$\,pc, \citealp{Gaia2018}) and youngest low-mass stars with a debris disk. Its spectral energy distribution (SED) shows a clear infrared (IR) excess already at {\it Spitzer}/IRS wavelengths and detections in the far-IR and sub-mm. On multiple occasions, the shape of the IR excess has been interpreted as the presence of two spatially separated dust belts, making TWA\,7 a very interesting system, with a possible gap. In Cycle\,3, we were awarded ALMA time to spatially resolve the dust distribution at sub-mm wavelengths. We present in this paper the results of those observations. 

\section{Observations, data reduction, and results}

\subsection{ALMA observations}

\begin{table*}
\centering
\caption{Log for the ALMA observations of TWA\,7, Prog ID. 2015.1.01015.S\label{tab:log}}
\begin{tabular}{@{}lclccccc@{}}
\hline\hline
Observing date     & Mode & Spectral Windows & Bandwidth & PWV & Beam size  & Integration & Maximum Recoverable Scale \\
$[$YYYY-MM-DD$]$   &      & $[$GHz$]$  &  $[$GHz$]$     & $[$mm$]$ & $[\arcsec]$ & $[$min$]$ & $[\arcsec]$\\
\hline
\multicolumn{8}{c}{Compact configuration - Pointed at $10$h$42$m$29.9$s $-33^{\circ}40^{\prime}17.1^{\prime\prime}$ - $42$ antennas} \\
\hline
\multirow{2}{*}{2016-04-22}  & Continuum & $334.352$, $336.242$, $348.35$  &   $2.0$ & \multirow{2}{*}{0.9} & \multirow{2}{*}{0.38} & \multirow{2}{*}{85.45} & \multirow{2}{*}{3.908} \\
             & Gas       & $345.950$ & $0.938$ & & & \\
\hline
\multicolumn{8}{c}{Extended configuration - Pointed at $10$h$42$m$30.4$s $-33^{\circ}40^{\prime}16.7^{\prime\prime}$ - $36$ antennas} \\
\hline
\multirow{2}{*}{2016-09-01} & Continuum & $334.352$, $336.242$, $348.35$  &   $2.0$ & \multirow{2}{*}{0.9} & \multirow{2}{*}{0.1} & \multirow{2}{*}{78.77} & \multirow{2}{*}{2.046} \\
             & Gas       & $345.950$ & $0.938$ & & & \\
\multirow{2}{*}{2016-09-15} & Continuum & $334.352$, $336.242$, $348.35$  &   $2.0$ & \multirow{2}{*}{0.9} & \multirow{2}{*}{0.1} & \multirow{2}{*}{81.29} & \multirow{2}{*}{2.046} \\
             & Gas       & $345.950$ & $0.938$ & & & \\
\hline
\end{tabular}
\end{table*}

\begin{figure}
\centering
\includegraphics[width=\hsize]{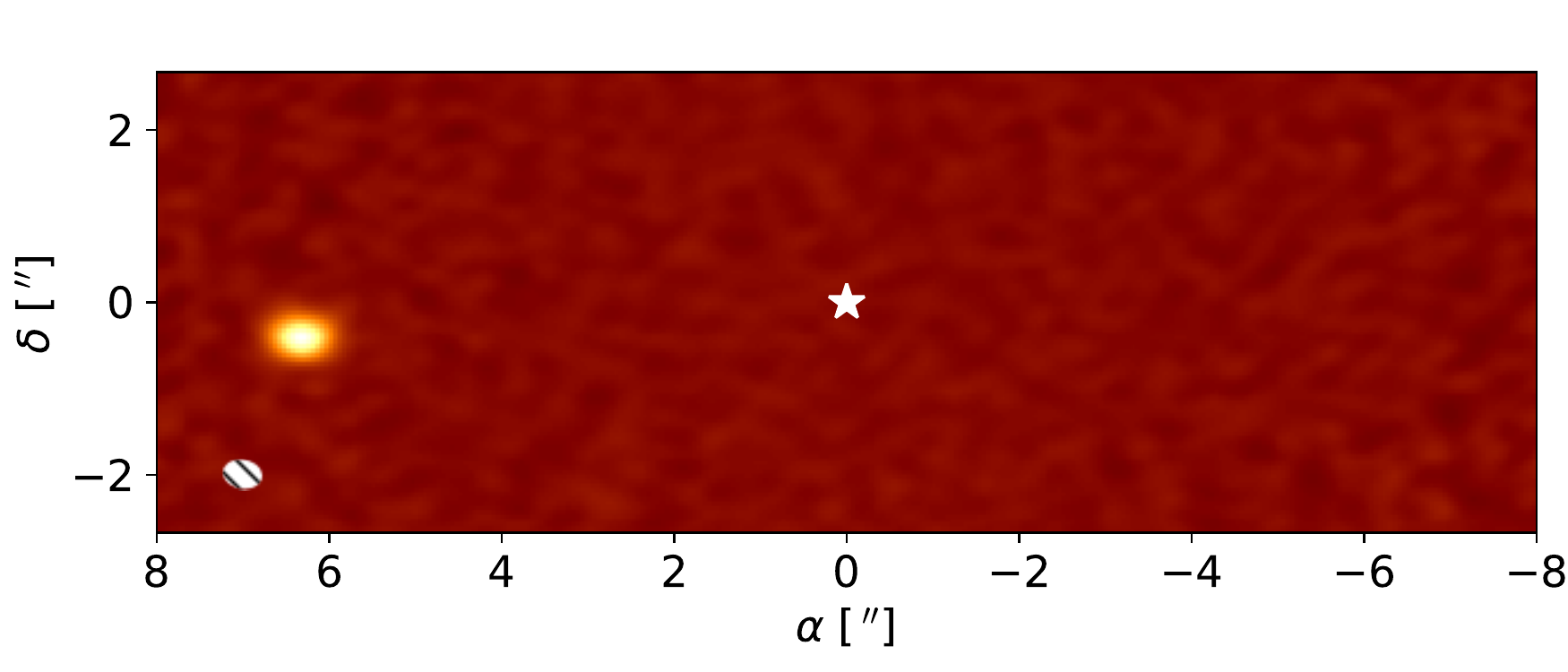}
\caption{Primary beam corrected image of the 1st ALMA Band\,7 observations. The position of TWA\,7 is marked with a white star. The beam size is shown in the bottom left corner.}
\label{fig:alma}
\end{figure}

TWA\,7 was observed twice with ALMA (project 2015.1.01015.S, PI: Bayo), in Band\,7 (346 GHz), using two different configurations (see Table\,\ref{tab:log} for a summary). The first configuration, observed on the 2016-04-22, the compact one, yielded a beam size of $0.44\arcsec \times 0.33\arcsec$. The second dataset, obtained on the 2016-09-01, led to a smaller beam size of $0.11\arcsec \times 0.09\arcsec$. Both datasets were reduced using CASA (version 4.7, \citealp{McMullin2007}), where Briggs weighting was selected by default in the CLEAN algorithm. Figure\,\ref{fig:alma} shows the continuum primary beam corrected image for the first dataset, with a total aggregated bandwidth of $6.938$\,GHz. The expected location of TWA\,7 is marked with a white star symbol but no emission is detected (see Sec.~\ref{sec:detection} for the results of a revised weighting of the visibilities). On the other hand, a compact source is detected at about $\sim 6.6\arcsec$ East of TWA\,7 ($\sim$2.95 mJy peak flux, at coordinates $10$h$42$m$30.42$s $-33^{\circ}40^{\prime}16.69^{\prime\prime}$). At the center of the image, we measure a root mean square of $32$\,$\mu$Jy.beam$^{-1}$. The analysis of the detected CO line is presented in \citet{Matra2019}.

Given that the pointing accuracy of ALMA cannot explain the $6.6\arcsec$ offset, we initially thought that there had been either an offset in the Scheduling Block, or that the proper motion (ppm) of TWA\,7 was erroneous. Therefore, for the second dataset, we modified the coordinates in the Scheduling Block, to have the compact source at the center of the image (and be able to obtain a more reliable flux estimate). With those observations (not shown here), we detect the same compact source, at the same coordinates. In addition, unfortunately, this change in the pointing led the current position of TWA\, 7 (taking into account the literature ppm and the work presented in this paper) to be outside of the recoverable field of view of the second data-set.

There are therefore two possible explanations: either the ppm of TWA\,7 is incorrect, or we detected something else than the debris disk and the signal from the latter has been filtered out even in the compact configuration (see Sec.~\ref{sec:detection}).

\subsection{Verifying the proper motion of TWA7}

\begin{figure}	
\centering
\includegraphics[width=\hsize]{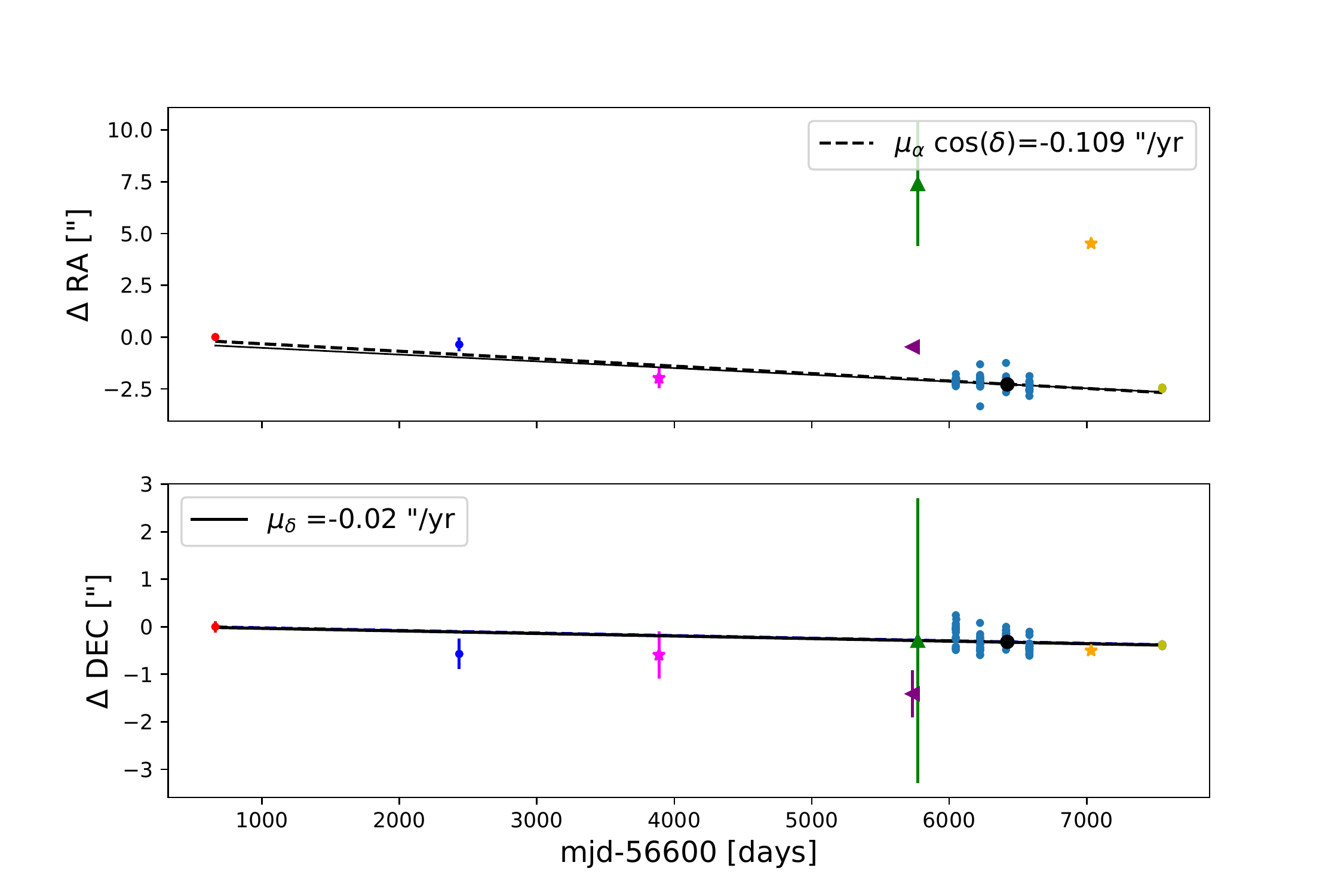}
\caption{Relative positions (with respect to 2MASS) in RA and DEC (upper and lower panel, respectively) for TWA\, 7 as a function of time.
The colors correspond to different observations: red 2MASS, cyan MIPS Spitzer/MIPS (24$\mu m$), magenta Spitzer/IRS (CASSIS program), purple Herschel , green JCMT, blue WISE and NEO WISE, black GAIA DR1, orange ALMA, yellow optical g and i bands. The solid and dashed black lines show the GaiaDR2 and our estimates of the ppm in each coordinate.}
\label{fig:PM}
\end{figure}

Given the discrepancy between the position of the ALMA detection and the expected position from the ppm and parallax for TWA\,7, we decided to search as many archives as possible for positions at different epochs  and acquire new observations in the optical for this source to study its ppm.
We retrieved positions from the following data: optical from EFOSC at NTT telescope in La Silla, Chile \citep{Ducourant2014};  the Gaia DR1 \citep{GaiaDR12016}; and a new dataset of g, r, i band images from the PUC observatory (a $40$\,cm telescope located in Santiago, Chile); near-IR 2MASS \citep{Skrutskie2006}, mid-IR Spitzer IRS (CASSIS); MIPS at 24$\mu m$, WISE and NEOWISE \citep{Lebouteiller2011,Low2004,Wright2010,Mainzer2014}; far-IR Herschel \citep{Riviere-Marichalar2013} and radio JCMT \citep{Holland2017}, and ALMA (this paper). The positions from Herschel, JCMT and ALMA were ignored in the final astrometric fitting process, see discussion in the following sections.
In Fig.\,\ref{fig:PM} we show the observed positions as a function of time highlighting with different colors, observations from different instruments/wavelengths and the best linear fit to the motion. The ppms from Gaia DR2 and our results are  displayed as a black solid and dashed lines, respectively. Our estimates for the ppm are, $\mu_ \alpha$ cos$(\delta)=-109 \pm 1$\,mas.yr$^{-1}$ and $\mu_\delta=-20 \pm 1$\,mas.yr$^{-1}$; consistent with previous values available in the literature 
\citep{Ducourant2014} 
and with the now available Gaia DR2 ($\mu_{\alpha _{Gaia}} = -118.934 \pm 0.114$\,mas.yr$^{-1}$, $\mu_{\delta _{Gaia}}$= $-19.851 \pm 0.106$\,mas.yr$^{-1}$, $\pi_{ Gaia} = 29.3867 \pm 0.0658$\,mas; $d$ $= 34.02 \pm 0.07$\,pc.). It is important to notice that we did not attempt to fit for parallax motion, and \citet{Ducourant2014,Gaia2018} do include that extra term.

\subsection{Ancillary far-IR and sub-mm observations}

It is not uncommon for the sub-mm emission from a debris disk to be (partially or fully) contaminated by a background galaxy (e.g., the unfortunate alignment for HD\,95086, \citealp{Su2017}). As debris disks' emission usually starts to become significant at far-IR wavelengths, with spatially unresolved observations, it is challenging to estimate whether the emission indeed arises from the system or not. 

\begin{figure}
\centering
\includegraphics[width=\hsize]{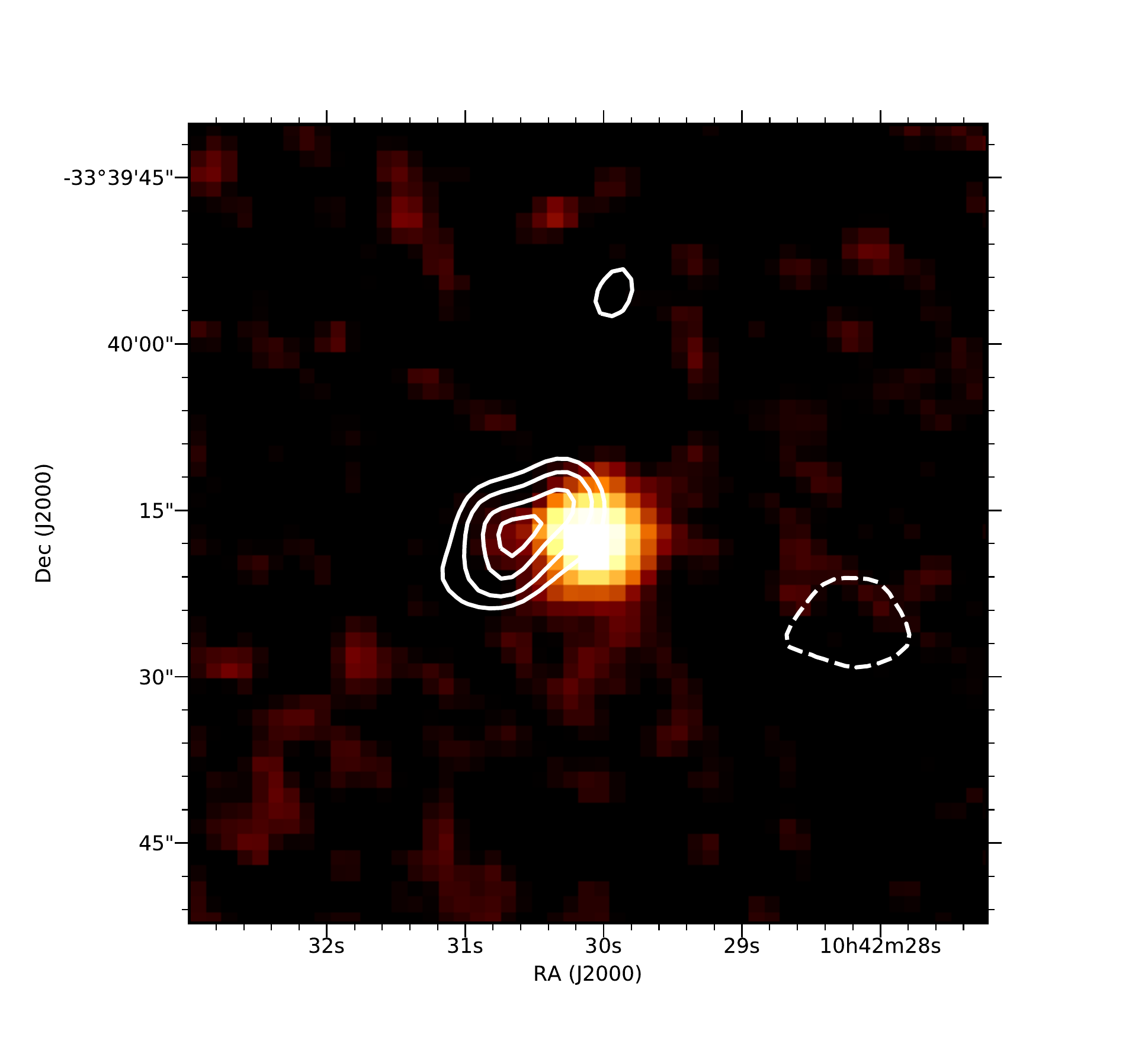}
\caption{\textit{Herschel}/PACS map at $100$\,$\mu$m, with the \textit{JCMT}/SCUBA-2 $850$\,$\mu$m contours overlaid.}
\label{fig:jcmt}
\end{figure}

In the case of TWA\,7, excess emission associated with a disk has been reported in \citet[][$70$\,$\mu$m, \textit{Spitzer}/MIPS]{Low2005}, \citet[][$70$, $100$, and $160$\,$\mu$m, \textit{Herschel}/PACS]{Riviere-Marichalar2013}, \citet[][$70$, $160$, $250$, and $350$\,$\mu$m, \textit{Herschel}/PACS \& SPIRE]{Cieza2013}, \citet[][$450$ and $850$\,$\mu$m, JCMT/SCUBA]{Matthews2007}, and \citet[][$850$\,$\mu$m, JCMT/SCUBA-2]{Holland2017}. In \citet{Holland2017}, the authors also reported a $\sim 6$\arcsec offset with respect to the expected position of TWA\,7, but could not firmly conclude on the nature of the sub-mm emission. Figure\,\ref{fig:jcmt} shows the \textit{Herschel}/PACS observation at $100$\,$\mu$m and the SCUBA-2 $850$\,$\mu$m observations as contours. While the location of TWA\,7 coincides with the position of the PACS source, the offset becomes clearly visible as the wavelength increases. Note that the SCUBA-2 data was obtained in 2013, while the Herschel data dates from 2010 and the ppm of TWA\,7 would translate in a smaller offset in the opposite direction. Therefore, it seems that the background contaminant does not contribute significantly at \textit{Herschel}/PACS wavelengths, and that those data points are indeed tracing the thermal emission of dust grains in the debris disk. The situation becomes less clear for the \textit{Herschel}/SPIRE observations published in \citet{Cieza2013}, especially the $350$\,$\mu$m, as the far-IR slope seems to become shallower (see Fig.\,\ref{fig:sed}). The beam size of SPIRE being $17.9$ and $24.2\arcsec$ at $250$ and $350$\,$\mu$m, respectively, we cannot assess whether the background object is contributing significantly at those wavelengths. 

\subsection{Filtering in the Fourier space: recovering emission from the disk}
\label{sec:detection}

The visibilities at the combination of baselines from the ALMA data, imaged with Briggs weighting (Fig.~\ref{fig:alma}), did not reveal any emission at the now secured position of TWA\,7. However, motivated by the extended nature of the disk in the NICMOS and SPHERE images ($\sim 2\arcsec$, \citealt{Choquet2016,Olofsson2018}), we tested re-imaging of the visibilities using different baseline weighting schemes. First, we combined both the compact and extended configurations together, as the position of TWA\,7 remains in the field of view of the second observations, despite the pointing offset. We then applied a $2\arcsec$ u-v taper to down-weight long baselines in favor of shorter baselines where most of the belt's emission may lie. This reduces our sensitivity to compact structure on scales less than $2\arcsec$ but allows us to recover any structure from larger scales that would otherwise be interferometrically filtered out of the image.

After application of this u-v taper, as well as the emission from the contaminating source, we also recover extended emission from the belt centered at the expected position of TWA 7. The emission is detected above the $4\sigma$ level and likely marginally resolved at the $\sim 3\arcsec$ resolution attained with u-v tapering.

To confirm that this extended emission component at the location of the star is not an artifact of imaging given the nearby, bright contaminant source, we model and then subtract the latter from the visibilities in u-v space. Since the contaminant is resolved in the Briggs-weighted image, we model it as a 2D Gaussian in u-v space using the \texttt{uvmodelfit} task within CASA, finding a best-fit flux of $3.9$\,mJy, a FWHM of $0.27\arcsec$,  an inclination of $\sim 45^{\circ}$ degrees from face-on and a position angle of $\sim 106^{\circ}$.

\begin{figure}
\centering
\includegraphics[width=\hsize]{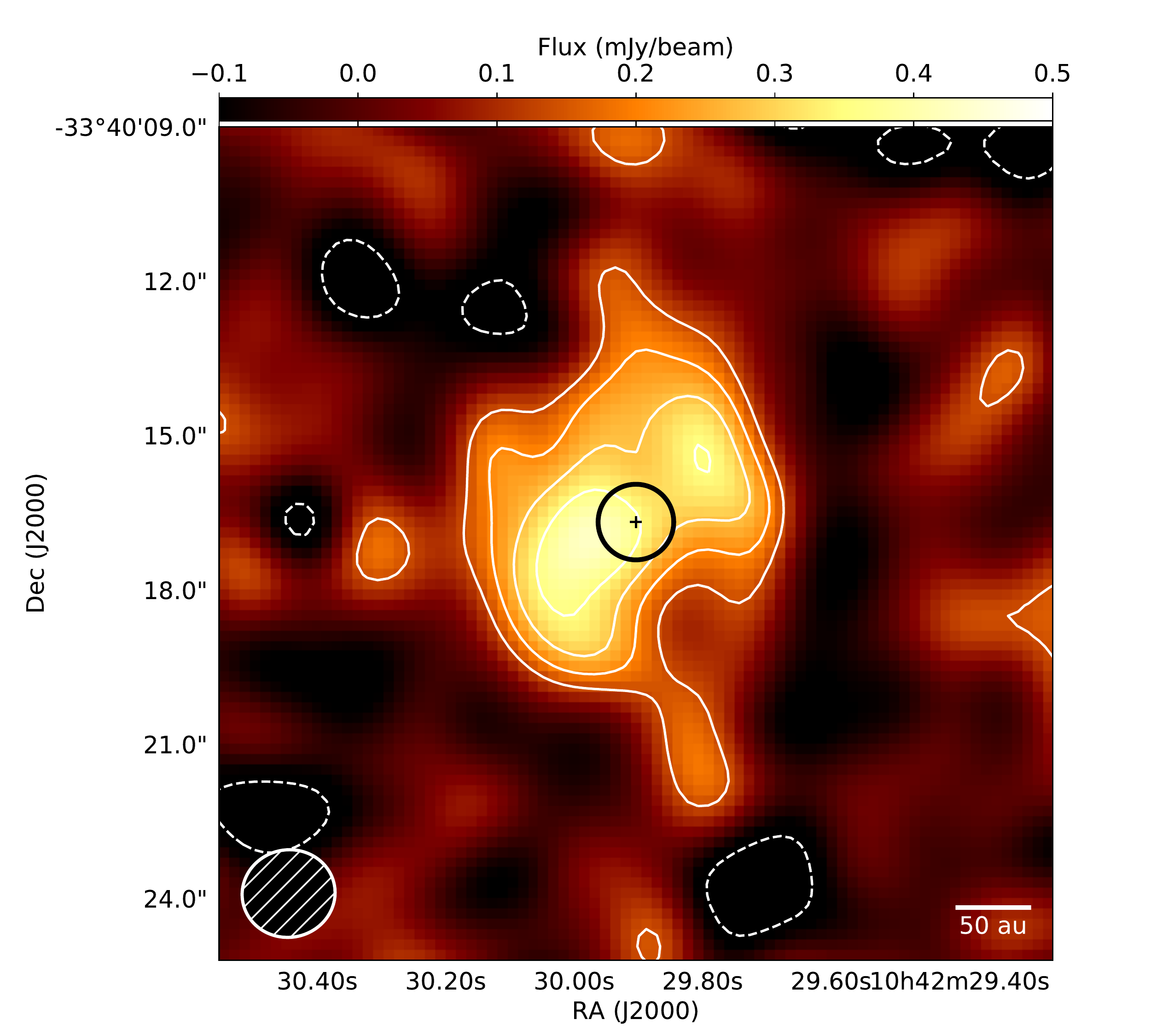}
\caption{Reconstructed image combining both configurations, after u-v taper ($3\arcsec$, see text for details) and removing the compact source. The contours enclose regions $[-2$,$2$,$3$,$4$,$5]$ times the RMS in the uv-tapered images ($70$\,$\mu$Jy / beam). The black circle show the peak position of the disk, as inferred from SPHERE observations (\citealp{Olofsson2018}).}
\label{fig:detection}
\end{figure}

Fig.\,\ref{fig:detection} shows the imaged visibilities of the contaminant-subtracted dataset, once again using a $2\arcsec$ taper. The belt is clearly detected with an integrated flux (measured by integrating emission in the region where the belt is detected at the $>2\sigma$ level) of $2.1 \pm 0.4$\,mJy (including the statistical uncertainty and a $10$\% flux calibration uncertainty added in quadrature). One should note that both TWA\,7 and the contaminant source have comparable flux densities (a few mJy). However, the contaminant appears much brighter in the left panel of Fig.\,\ref{fig:detection} because it is much more compact while the debris disk is more extended and therefore the total flux is more diluted. 

\section{Revised SED model}

Knowing that some of the far-IR and sub-mm observations are most likely dominated by the background source that we identified with our ALMA dataset, and with our new detection in hand, it is necessary to revisit the SED modeling of TWA\,7, especially with respect to the number of dust belts that has been inferred for the disk on many occasions (\citealp{Matthews2007}; \citealp{Riviere-Marichalar2013}; \citealp{Holland2017}; \citealp{Olofsson2018}). We refer the reader to \citet{Olofsson2018} for the description of the stellar properties and of the code used to model the SED. The only difference is that we now treat the JCMT reported fluxes as non-detections, and include the new ALMA point in the modeling. Concerning the far-IR points, we chose to include the $250$\,$\mu$m \textit{Herschel}/SPIRE point as it appears consistent with the slope between the $160$\,$\mu$m and ALMA points, but we did not include the $350$\,$\mu$m observation. One has to note that there are some marginal discrepancies between the reported fluxes at $70$ and $160$\,$\mu$m between \citet{Riviere-Marichalar2013} and \citet{Cieza2013} ($77 \pm 7$ and $42 \pm 9$\,mJy at $70$ and $160$\,$\mu$m for \citealp{Riviere-Marichalar2013}, to be compared with $68.8 \pm 3.13$ and $49.8 \pm 7.05$\,mJy at the same wavelengths for \citealp{Cieza2013}). Despite those small differences, the fluxes remain consistent with each other within their corresponding uncertainties, and we opted to use all of them. For the JCMT/SCUBA and SCUBA-2 non-detections, the $3\sigma$ upper limits are estimated from the uncertainties reported by \citet[][at $450$\,$\mu$m]{Matthews2007} and \citet[][at $850$\,$\mu$m]{Holland2017}. 

To alleviate some of the known degeneracies of SED modeling of debris disks (mostly the distance to the star and the minimum grain size), we reduce the number of free parameters to the strict minimum. We consider that there is only one dust belt around the star, and based on the analysis of the SPHERE images presented in \citet{Olofsson2018}, we fix its reference radius $r_0 = 25.0$\,au (consistent with the marginally resolved ALMA emission) and the inner slope of the dust density distribution $\alpha_{\mathrm{in}} = 5.0$, with the dust density distribution following
\begin{equation}\label{eqn:nr}
n(r) \propto \left[\left(\frac{r}{r_0}\right)^{-2\alpha_{\mathrm{in}}} + \left(\frac{r}{r_0}\right)^{-2\alpha_{\mathrm{out}}}\right]^{-1/2}
\end{equation}
This leaves as free parameters the outer slope $\alpha_{\mathrm{out}}$, the minimum grain size $s_{\mathrm{min}}$ and the dust composition. When computing a model for a given set of parameters, the dust mass is not an input, but it is evaluated automatically by scaling the model to best reproduce the observations, using a least squares method. The grain size distribution follows a differential power-law d$n(s) \propto s^p$d$s$ (with $p$ fixed to $-3.5$ following \citealp{Dohnanyi1969}, with fixed $s_{\mathrm{max}} = 5$\,mm). 

Concerning the dust composition, we used the astro-silicates optical constant from \citet[][density of $3.5$\,g.cm$^{-3}$]{Draine2003}, similarly to \citet{Olofsson2018}, but one has to keep in mind that using different dust composition would most likely lead to different temperature distribution as a function of the grain size. To find the most probable solution, we used the affine invariant ensemble sampler (\texttt{emcee}, \citealp{Foreman-Mackey2012}), with $20$ ``walkers'', a burn-in phase of $1,000$ steps, and a final chain of $2,000$ steps. We obtained an acceptance ratio of $0.44$, and auto-correlation lengths of $81$. 

\begin{table}
\centering
\caption{Best fit results for the modeling of the SED.\label{tab:sed}}
\begin{tabular}{@{}lccc@{}}
\hline\hline
Parameter               & Uniform prior  & $\sigma_{\mathrm{kde}}$ & Best-fit value \\
\hline
$\alpha_{\mathrm{out}}$          & $[-30, -0.5]$ &  $0.1$    & $-4.5_{-0.3}^{+0.2}$ \\
$s_{\mathrm{min}}$ [$\mu$m]      & $[0.01, 5.0]$ &  $0.01$   & $0.16_{-0.02}^{+0.04}$ \\
M$_{\mathrm{dust}}$ [M$_\oplus$] &               &  $0.0005$  & $0.018_{-0.001}^{+0.002}$ \\
\hline
\end{tabular}
\end{table}

\begin{figure}
\centering
\includegraphics[width=\hsize]{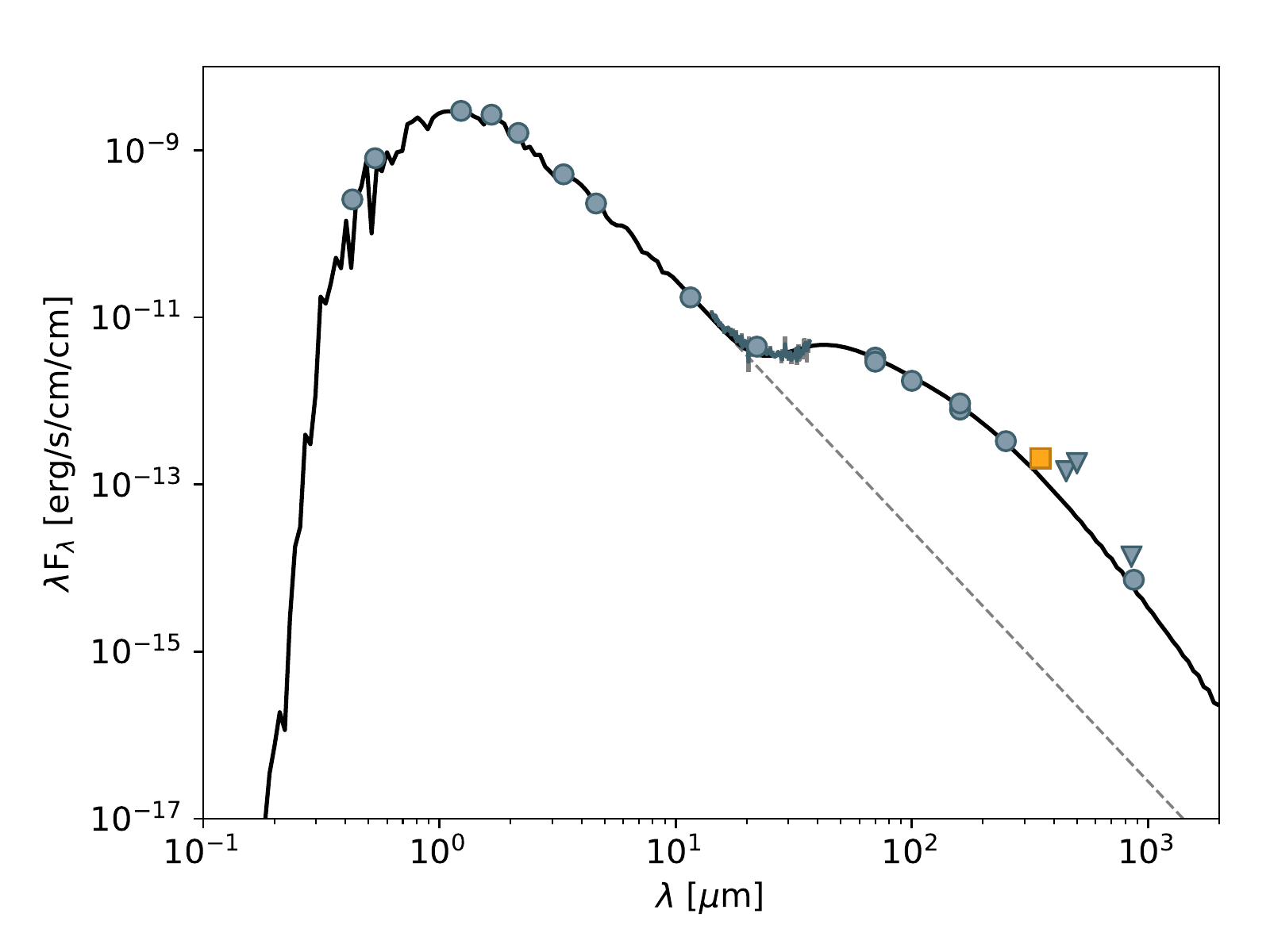}
\caption{Results of the SED modeling, assuming astro-silicate optical properties. The $350$\,$\mu$m \textit{Herschel}/SPIRE point (possibly contaminated datapoint) is shown as an orange square, for display purposes, and the downward triangles represent upper limits.}
\label{fig:sed}
\end{figure}

From the projected probability distributions, we estimated the best fit parameters as well as their respective uncertainties (using a kernel density estimation, with width $\sigma_{\mathrm{kde}}$), which are reported in Table\,\ref{tab:sed}. The best-fit model is shown in Figure\,\ref{fig:sed} (where the possibly contaminated \textit{Herschel}/SPIRE point is shown in orange, and upper limits are shown as downwards triangles). First of all, it is clear that the SED can be reproduced without invoking any additional belt, neither inward nor outward of the belt seen with HST and SPHERE. 

We find that the minimum grain size is of $\sim 0.16$\,$\mu$m, and the volumetric dust density distribution has a slope in $\sim -4.5$ (equivalent to a surface density of $\sim -3.5$). For comparison, \citet{Olofsson2018} modeled the SPHERE image with a very extended disk ($\alpha_{\mathrm{out}} \sim -1.5$), but the comparison with results from SED modeling is not straightforward. High-contrast images reveal the population of very small dust grains, which are very sensitive to radiation pressure (and stellar winds for low-mass stars), while with the far-IR thermal emission we are sensitive to a wider range of grain sizes. Therefore, scattered light images and the SED do not necessarily probe the same populations of dust grains. 

Similarly to \citet{Olofsson2018}, we estimated the average stellar wind speed that would explain the minimum grain size inferred from SED modeling ($s_{\mathrm{min}} \sim 0.16$\,$\mu$m). We computed the $\beta$ ratio between radiation pressure and gravitational forces, as well as the $\beta_{\mathrm{wind}}$ exerted by stellar winds onto the small dust grains
\begin{equation}
\beta_{\mathrm{wind}} = \frac{3}{32\pi}\frac{\dot{M}_\star v_\mathrm{sw}C_\mathrm{D}}{G M_\star \rho s},
\end{equation}
where $\dot{M}_\star$ is the stellar mass loss rate, $C_\mathrm{D}$ a factor close to $2$, $v_\mathrm{sw}$ the speed of the stellar wind (assumed to be $400$\,km.s$^{-1}$), $G$ the gravitational constant, and $\rho$ the dust density. We find that $\beta + \beta_{\mathrm{wind}} = 0.5$ can be reached for a size of $\sim 0.16$\,$\mu$m for $\dot{M}_\star \sim 550 \dot{M}_\odot$. As a comparison, \citet{Augereau2006} found an average stellar mass loss rate of about $300$\,$\dot{M}_\odot$ for AU\,Mic (assuming flares are present $10$\% of the time). Given the fact that TWA\,7 is younger than AU\,Mic, and therefore more active, the comparison does not appear incompatible if flares would happen $\sim 18$\% of the time.

\section{Conclusions}

In this paper we presented our ALMA observations pointed toward the debris disk around the low-mass star TWA\,7. We confirmed the shift of $6.6\arcsec$ for the peak of the sub-mm emission that was initially reported in \citet{Holland2017}, and demonstrate unambiguously that a background source dominates the sub-mm flux. The disk around TWA\,7 is nonetheless detected, at its expected position, with an integrated flux density of $2.1$\,mJy at $870$\,$\mu$m.

We presented a revised model for the SED, and concluded that IR emission can be well reproduced assuming a single dust belt (the one detected in scattered and polarized light with HST and SPHERE). With the new ALMA detection and the revised geometry of the system, we find that the total dust mass (with sizes smaller than $5$\,mm) is of the order of $2 \times 10^{-2}$\,M$_\oplus$. Finally, we found that the revised value for $s_{\mathrm{min}}$ is roughly compatible with an active low-mass star.

\section*{Acknowledgements}
We thank W. Holland for sharing the SCUBA-2 data.
A.\,B., J.\,O., J.C.\,B., M.\,S., I.\,G., and C.\,C., acknowledge support from ICM (Iniciativa Cient\'ifica Milenio) via the N\'ucleo Milenio de Formaci\'on Planetaria. A.\,B. acknowledges support from FONDECYT (grant 1190748). J.\,O. acknowledges support from Universidad de Valpara\'iso, and FONDECYT (grant 1180395). J.C.\,B. acknowledges support from FONDECYT (grant 3180716). L.M. acknowledges support from the Smithsonian Institution as a SMA Fellow. M.B. acknowledges support from the Deutsche Forschungsgemeinschaft, project Kr 2164/15-1. C.C. acknowledges support from project CONICYT PAI/Concurso Nacional Insercion en la Academia, convocatoria 2015, folio 79150049. This paper makes use of the following ALMA data: ADS/JAO.ALMA\#2015.1.01015.S. ALMA is a partnership of ESO (representing its member states), NSF (USA) and NINS (Japan), together with NRC (Canada), MOST and ASIAA (Taiwan), and KASI (Republic of Korea), in cooperation with the Republic of Chile. The Joint ALMA Observatory is operated by ESO, AUI/NRAO and NAOJ.





\bibliographystyle{mnras}
\bsp	
\label{lastpage}
\end{document}